# The Ancient Astronomy of Easter Island: Venus and Aldebaran

Sergei Rjabchikov[1]

[1]The Sergei Rjabchikov Foundation - Research Centre for Studies of Ancient Civilisations and Cultures, Krasnodar, Russia, e-mail: srjabchikov@hotmail.com

**Abstract**

One additional position of the famous Mataveri calendar of Easter Island has been interpreted. New data on the watchings of Venus and Aldebaran have been rendered. Some reports about the sun, the moon as well as Sirius are of our interest, too.

**Introduction**

Popova (2015) has disclosed a very interesting *rongorongo* record concerning Venus and Aldebaran on the Great Santiago tablet (H). In this light, the author tries detecting new data about both celestial bodies in Rapanui sources.

**A Farther Remark about a Rapanui Rock Calendar**

On a boulder at Mataveri (a crucial area of bird-man rituals) some lines were incised; most of them were the directions of the setting sun according to Liller (1989). I have calculated the corresponding days for the year 1775 A.D. (Rjabchikov 2014: 5, table 2; 2015: 2, table 1). Here and everywhere else, I use the computer program RedShift Multimedia Astronomy (Maris Multimedia, San Rafael, USA) to look at the heavens above Easter Island. For the date of November 12 one can realise that ancient priests-astronomers observed the appearance of Venus before dawn.

 **Table 1.** The Dates Calculated (with the interpretation for November 12):

**June 22** (the azimuth of the sun = 296.2°): one day after the winter solstice;
July 21 (292.5°): the star Capella (α Aurigae) before dawn;
August 11 (286.7°): the star Pollux (β Geminorum) before dawn;
September 2 or 3 (277.9°): the star β Centauri [*Nga Vaka*] before dawn;
**September 21** (270.1°): the day before the vernal equinox, the key moment of the bird-man feast;
September 24 (268.7°): the new moon;
September 27 (267.4°); the fourth night: the measure of the visible dimensions of the moon;
October 1 (265.9°); the eighth night: the measure of the visible dimensions of the moon;
October 3 (264.7°);
October 22 (256.8°): near the new moon;
November 8 (250.7°): the star Spica (α Virginis) before dawn;
November 12 (249.3°); **Venus as the Morning Star before dawn**;
November 14 (248.7°);
November 23 (246.3°): the new moon;
**December 20** (the azimuth of Aldebaran = 339.1°): the star Aldebaran (α Tauri) at night;
**December 21** (the azimuth of Aldebaran = 322.1°; the azimuth of Canopus = 177.5°): the stars Aldebaran (α Tauri) and Canopus (α Carinae) on the same night (Rjabchikov 2013: 7); the day of the summer solstice.

# Again on the Record about Venus and Aldebaran

I would like to examine the whole *rongorongo* text associated with the chant "*E rua nga uka*" (Two girls), see figure 1. It was a record in a textbook from the school of king *Nga Ara*. Fundamentally, I use the drawings of the classical *rongorongo* inscriptions published by Barthel (1958).

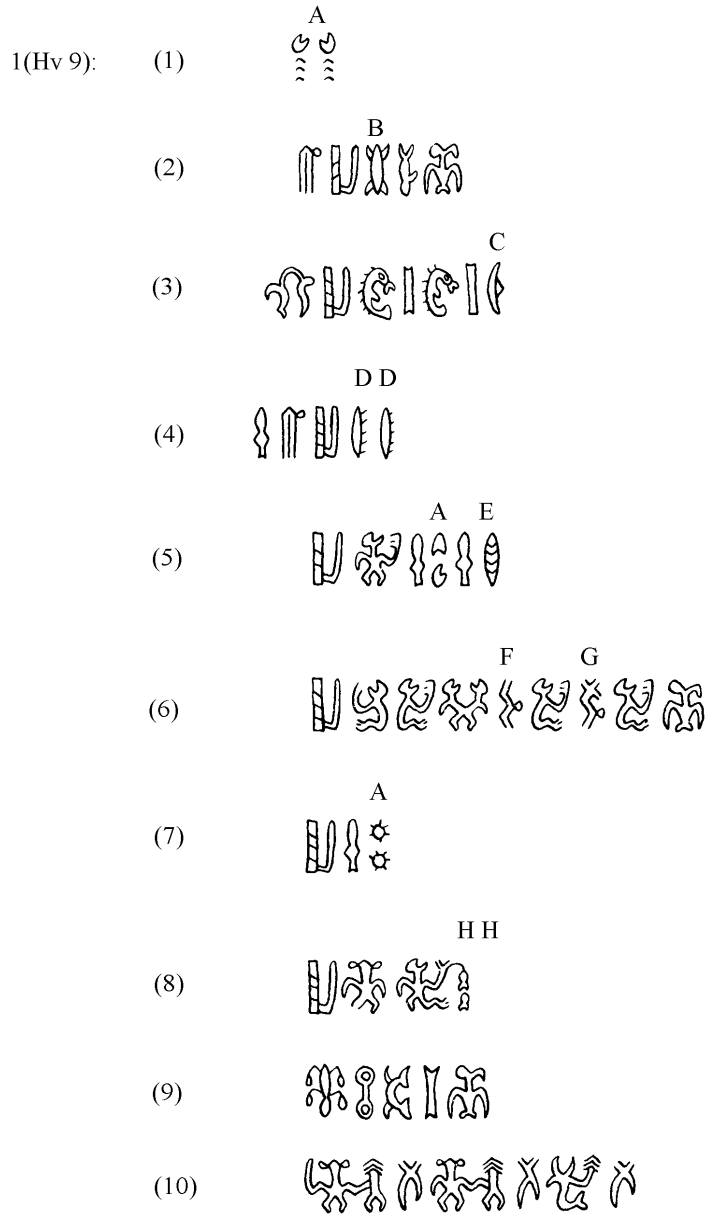

Figure 1

(1) **24-24 29var** *Aiai rua*. 'A host of the places (= the glyphs) *rua*.'
(2) **26 4-4 28 9 44** *Maa, atiati: ngaa – niva taha*. '(You) learnt, wrote: the break of day – the blackness of the horizon.'
(3) **44b 4-4 11var 4 11var 4 3** *Tua, atiati: poki – ati, poki – ati, hina*. '(You) spoke, wrote: a child – write, a child – write – WOMAN (*hahine*).' (Cf. Tuamotuan *tua* 'to speak'.)
(4) **73 26 4-4 46-46** *He maa, atiati: naanaa*. '(You) learnt, wrote: (It was) the disappearance.'
(5) **4-4 73 29 73 33** *Atiati: Ha e rua, e ua*. '(You) wrote: Two girls went in darkness.'

(6) **4-4 6-6 6 50 6 70 6-44** *Atiati: Haha, ha Hi, ha Pu, hata*. '(You) wrote: *Hi(a)* [Venus] went in darkness, *Pu(a)* [Aldebaran] went in darkness; (it) rose.'
(7) **4-4 73 29var** *Atiati: e rua*. '(You) wrote: the two (girls).'
(8) **4-4 6-6 12-12** *Atiati: Haha ika, ika*. '(You) wrote: two (celestial) bodies went in darkness.'
(9) **25 65 8 4-44** *Hua rangi, motu Tita(h)a*. 'The sons (= schoolmates) called (such words), (they) wrote at the Hare Tita(h)a.'
(10) **6 33-59 15-25 6 33-59 15-25 11 33 15-25** *Ha uka – rohu! Ha uka – rohu! Poki ua – rohu!* 'The (first) girl went in darkness – write! The (second) girl went in darkness – write! The children/girls – write!'

As a new version of the chant let us pick out these words on the Aruku-Kurenga tablet (B), see figure 2. This drawing has been slightly corrected.

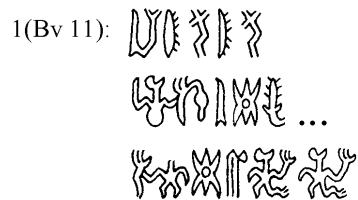

Figure 2.

1 (Bv 11): **5-15 46-70-46-70 68 15-25 5 7 25 … 15 99var 7 26 6-6** *Atua roa naa Pu, naa Pu. Honu rohu atua Tuu Hu(a)... Romi Tuu maa, haha…* '(It was) the great deity 'Aldebaran.' The Pleiades created (= appeared before) the deity (the star) 'Aldebaran'… The bright star (Aldebaran) was covered (with rainy clouds), it went in darkness.'

When I visited the General Archives of the Congregation of the Sacred Hearts of Jesus and Mary (Rome) in 2015, I carefully studied the signs on that tablet. The second glyph in the third segment is a human figure without the "eyes" = without the head. So, it is a variant of glyph **99** *mi*. A hole on the "neck" of this sign can be a score made by king Nga Ara himself with a shark's tooth. Remember that he organised special readings of *rongorongo* inscriptions and corrected mistakes of the orators (Routledge 1998: 245-246).

Thus, the interpretations of the terms *Pu(a)* and *Romi* for Aldebaran offered by Popova are true. In the Polynesian archaic astronomy Tahitian *Ana-muri* ['*Ana-muri*] 'the pillar to blacken or tattoo by' applied for the Aldebaran's designation (Henry 1907: 102) is a possible lexical parallel. Tahitian *muri* (behind) is a symbolic meaning of the shade and disappearance. The name **7 25** *Tuu Hu(a)* 'Aldebaran' (Rjabchikov 1993a: 6) signifies 'The star of the water,' cf. Samoan *sua* 'liquid' and Tongan *hū* 'wet.'

Furthermore, Old Rapanui *Tuu Hi(a)* 'The Star of the Sunbeams' (Venus as the Morning Star) can be compared, in my opinion, with Marquesan *Hetu ao* 'star-of-dawn' or 'Morning Star' (Makemson 1941: 207; Handy 1923: 352).

**On Constant Watchings at the Royal Observatory**

According to Mulloy (1973), the ceremonial platform Ahu Huri a Urenga was a real solar observatory. I have culled a report about this scientific centre in the record on the Aruku-Kurenga tablet, see figure 3.

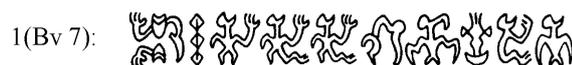

Figure 3.

1 (Bv 7): **53 29 102 17 6 6-6 44b 44 3 56-33 6-44** *Maru (Maro) Rua Ure tea. Ha haha: tua, ta(h)a hina, Pou hata…* '(It was) the month Maru (Maro; June chiefly). The observations (*tea*) were carried out at Rua Ure [Huri (= Uri) a Ure-nga]. (They were) the observations in darkness: the moon set, (and) Sirius rose.'

An important point is that the monument standing on that platform has two pairs of hands. I believe that the surplus hands denote glyph **53** *Maru* (*Maro*; the month of the winter solstice; the season of rains; it began in the new moon of June as a rule).

**On the Observations of the Moon and Venus in June**

Consider the following parallel records on the Great St. Petersburg (P), Great Santiago, Small St. Petersburg (Q) and Aruku-Kurenga tablets, see figure 4.

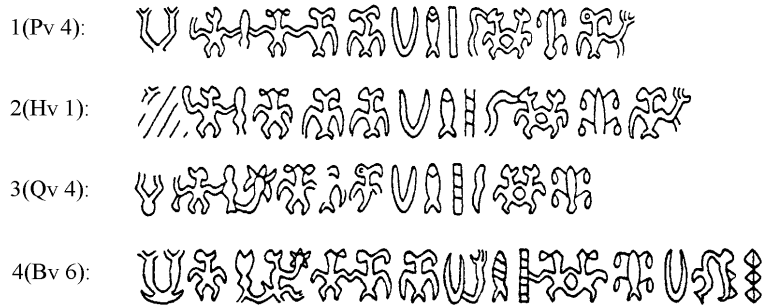

Figure 4.

1 (Pv 4): **53 6 73 6-44 44 61 12 4-50 6-25 44-15** *Maru (Maro) ha. He hata, ta(h)a Hina Ika, Tuhi(a) ahu Ta(h)a roa.* '(The moon of the month) *Maru (Maro)* (June chiefly) was in darkness (invisible). The moon (the fish) (connected) with Venus (*Tuhi = Tuu Hi*) rose (and) set at the ceremonial platform Taha roa.'
2 (Hv 1): **53 6 73 6-44 44 61 12 4 50 6-25 44-15** *Maru (Maro) ha. He hata, ta(h)a Hina Ika, Tuhi(a) ahu Ta(h)a roa.* '(The moon of the month) *Maru (Maro)* (June chiefly) was in darkness. The moon (the fish) (connected) with Venus (*Tuhi = Tuu Hi*) rose (and) set at the ceremonial platform Taha roa.'
3 (Qv 4): **53 6 73 7 6-44 44 61 12 4 50 6-25 6** *Maru (Maro) ha. He tuu, hata, ta(h)a Hina Ika,Tuhi(a) ahu.* '(The moon of the month) *Maru (Maro)* (June chiefly) was in darkness. The moon (the fish) (connected) with Venus (*Tuhi = Tuu Hi*) came, rose (and) set at a platform.'
4 (Bv 6): **3 53 6 73 6 7 6-44 44 61 15 12 4 6-25 61 44b-17** *Hina Maru (Maro) ha. He ha, tuu, hata, ta(h)a Hina Ika Atua. Ahu. Hina, Tua tea.* 'The moon (of the month) *Maru (Maro)* (June chiefly) was in darkness. The great moon (the fish-the goddess) went in the darkness, came, rose (and) set. (It was a certain) platform. (The conjunction of) the moon (and) Venus (occurred).'
    The conjunctions of the moon and Venus occurred, for instance,
    (1) on June 23, 1672 A.D., at 05:43 (the moon's azimuth: 59°03'43"; the rising sun: 07:07, its azimuth: 63°51'09");
    (2) on June 26, 1715 A.D., at 04:54 (the moon's azimuth: 65°46'31"; the rising sun: 07:07, its azimuth: 76°54'55");
    (3) on June 26, 1726 A.D., at 04:17 (the moon's azimuth: 63°41'54"; the rising sun: 07:07, its azimuth: 63°57'41").
    So, the rising moon and sun were seen in some cases from the environs of the platform Huri a Urenga in the direction of the platform Taha roa (the symbolism of the winter solstice?). In compliance with a Rapanui myth, the moon goddess (*Nuahine = Hina*) lived at the bay Hanga Taha roa on the northeastern coast of the island (Felbermayer 1973: 79-84).

**A New *Rongorongo* Record about Aldebaran and Venus**

Wieczorek and Horley (2015: 138, figure 7) have published the inscriptions on both sides of a replica of the so-called Lateran tablet. I presume that I merely heard about that original tablet from Fr. Lejeune in Rome on May 13, 2015. Plainly it belongs to the Congregation of the Sacred Hearts of Jesus and Mary (Rome), but now it is housed in the Vatican. Unfortunately, I lack all detailed information.

I have all the grounds to insist that this artefact was a late copy (without the reverse boustrophedon style mostly) of a real textbook for the study of the cursive version of the *rongorongo*. A new synopsis for the learning of the chant "*E rua nga uka*" and the instructions for the basic glyphs are presented below, see figure 5 (my own drawing).

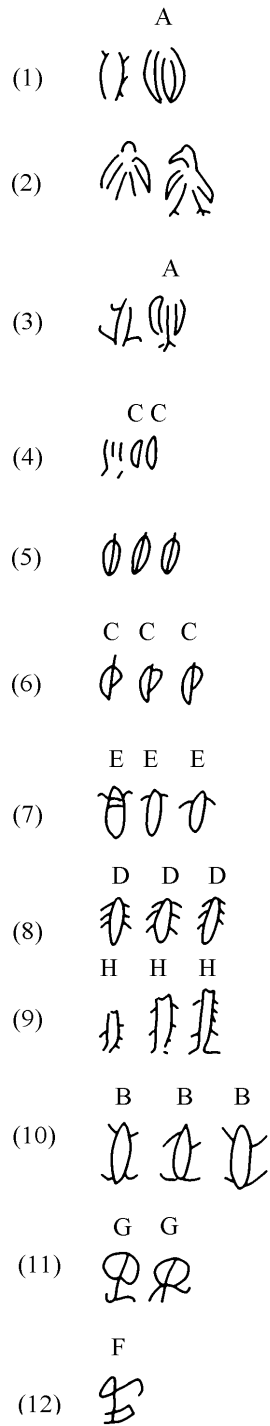

Figure 5.

(1) **28 29var** *Nga rua*! '(Write) a host of (the glyphs) *rua*!'

(2) **44-44** *Tahataha*! 'Turn (the tablet)!'
(3) **28 24 29var** *Nga ai rua.* 'The places (= the glyphs) *rua*.'
(4) **32 3 3** *Ua HINA, HINA.* 'Two girls.'
(5) **110 110 110** *Vie, vie, vie.* '(Repeat the glyphs) *vie* WOMAN.'
(6) **3 3 3** *Hina, hina, hina.* '(Repeat the glyphs) *hina* WOMAN.'
(7) **33 33 33** *Ua, ua, ua.* '(Repeat the glyphs) *u(k)a* GIRL.'
(8) **46 46 46** *Naa, naa, naa.* '(Repeat the glyphs) *naa* HIDDEN.'
(9) **12 12 12** *Ika, ika, ika.* '(Repeat the glyphs) *ika* BODY.'
(10) **28 28 28** *Ngaa, ngaa, ngaa.* '(Repeat the glyphs) *ngaa* (DAY)BREAK.'
(11) **70var 70var** *Pu, Pu.* 'Aldebaran.'
(12) **4-50** *Tuhi(a)* [*Tuu Hi(a)*]. 'Venus.'

**Cross-readings of Several Glyphs in the *Rongorongo***

1. Consider the following record on the Great Santiago tablet, see figure 6.

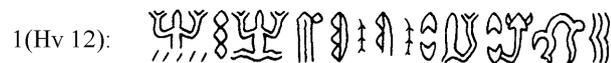

Figure 6.

1 (Hv 12): **69 17 69 26 18-24-18-24 29 48-15 29 21 44b 33** *Moko tea, moko maa, te ai, te ai: rua uri, rua kotua – ua.* 'The white lizard, the bright lizard (= the clear day sky), the same two places (glyphs) [*moko* 'lizard'] with the glyphs *rua uri* 'the black place of the sunset', with the glyphs *rua kotia* 'the sign *rua* (something cut in two)' or *rua ko tua* 'the sunset in west') – the rain.'

    A parallel segment is presented in the "*Apai*" chant: *mokomoko uri – ua; mokomoko tea* 'the black lizard (= the dark day sky) – the rain; the white lizard (= the clear day sky)' (Rjabchikov 1993b: 141).
    2. Consider the following record on the same tablet, see figure 7.

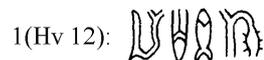

Figure 7.

1 (Hv 12): **5-15 1 12 15-25** *atua roa Tiki Ika rohu* 'the great deity *Tiki*, (a son of) the Fish (= the principal god *Tangaroa*)-the creator.'

    In Manuscript E (Barthel 1978: 304) there is such a passage: *Ko Tiki Hati a Tangaroa* '*Tiki Hati* (*Tiki te Hatu*), (a son of) *Tangaroa*.'
    3. Consider the following record on the same tablet, see figure 8.

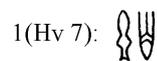

Figure 8.

1 (Hv 7): **73 1** *e Tiki* '*Tiki*.'
    The name of the sun god *Tiki* is introduced by the particle *e* for personal names.
    4. Consider the following record on the Aruku-Kurenga tablet, see figure 9.

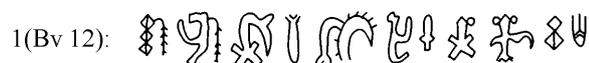

Figure 9.

1 (Bv 12): **17-24 62 24 44-41 9 15a** (it is the inverted sign) **14 6 73 49 6 67var 17 1** *Te ai To Arii: Tare, Niva roa hau ha, he mau. Api te Tiki.* '(It was) the place Tongariki [*To-nga Ariki*]: *Tare* (and) *Tive* (= *Hi-va*, *Niva* 'Blackness') created four (directional) winds (and) brought (the rain). *Tiki* hid himself.'

Here I have inserted my own reproduction of the inscription.

It is a version of a legend about the deity *Hiva Kara Rere*. According to that narration (Felbermayer 1971: 29-32), a priest once stood on the platform Tongariki and prayed for rain. He shouted: *Tare, e Tive, nga poki a Tiki, hakapiri te Tokerau o te papakina roua ko Poike! Ka hakapiri puhi Orongo ki Tongariki!* 'O *Tare* and *Tive*, the children of *Tiki*, gather the bad wind Tokerau and (the wind blowing) from Poike! Gather the wind (blowing) from Orongo and (the wind blowing) from Tongariki!' Then he shouted: *Tiki ka api tou aringa!* 'O *Tiki*, hide your face!' He shouted also: *E Tare, e Tive ka mau mai te rangi!* 'O *Tare* (and) *Tive*, bring the (rainy) clouds (sky)!' He said several incantations yet. As a result, the rain came.

So, *Tare* and *Tive* created four directional winds. The ghosts played the roles of east and west respectively (Rjabchikov 2001: 218). The key indicator of the second incantation is Rapanui *api* (to cover, to hide). The key indicator of the third incantation is Rapanui *mau* (to bring).

Undoubtedly, *Tiki* was the sun god. Let us read another folklore text (Felbermayer 1972: 275). This fragment sounds as follows: *E Tane, ka ata, hakaoho mai te rangi, mai Orongo, mai Poike! Ka ata: hea ua!* 'O *Tane*, hide himself, let (the winds) go off from the sky, from Orongo (and) from Poike! Hide himself for rain!' Hence, the East Polynesian god *Tane* as the sun deity once existed in the Rapanui pantheon.

5. Consider the following record on the Great Santiago tablet, see figure 10.

1(Hv 2): 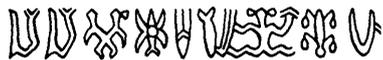

Figure 10.

1 (Hv 2): **5-15-5-15 6-7 1 27-32 69 33 25 61** *Atua roa, atua roa Hatu Tiki raua MOKO Vai hua hina* (= *marama*). 'The great deity, the great deity *Tiki-te-Hatu* (= *Tane*) together with *Vai* (Water) produced the moon.'

This variant of the creation myth correlates with a Maori one: *Tane*'s brother called *Tangotango* (*Tangaroa*?) and *Wainui* [*Wai nui* 'The great water'] produced *Te Marama* 'The moon' (Best 1922: 7).

**Conclusion**

I can say with confidence that the Rapanui archaeoastronomy became the main clue to the local script.

**Acknowledgements**

I wish to thank Fr. Paul Lejeune for his kind permission to study four excellent *rongorongo* tablets in the General Archives of the Congregation of the Sacred Hearts of Jesus and Mary (Rome) in May 2015. I am grateful to Ms. Luana Tarsi for her assistance during that research.

# Appendix

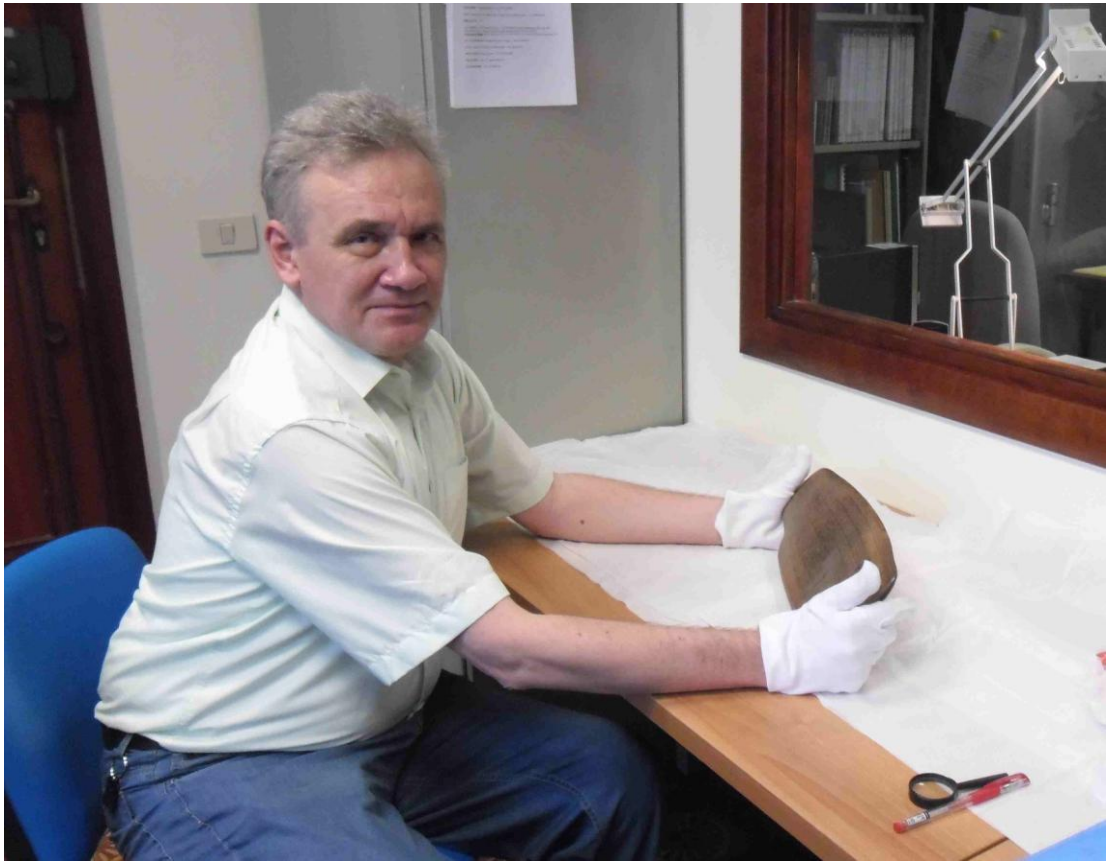

I visited the General Archives of the Congregation of
the Sacred Hearts of Jesus and Mary in 2015.
I was tracing the *rongorongo* glyphs of the Aruku-Kurenga tablet.

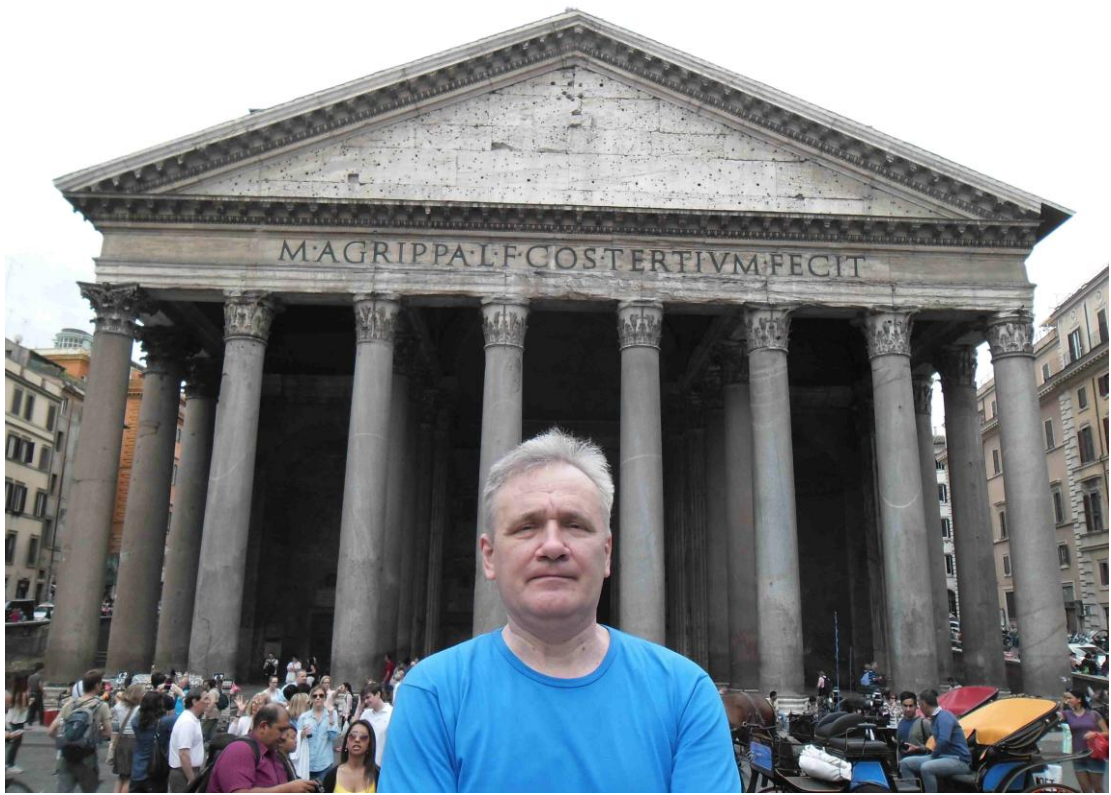

My attendance in Rome gave me an impetus to a new study of
the Rapanui records dedicated to zodiacal constellations.
I imagined that the Pantheon had been once decorated inside with zodiacal designs.